\documentclass[aps,prb,superscriptaddress,reprint,showpacs,twocolumn,preprintnumbers,floatfix]{revtex4-2}
\usepackage[dvips]{graphicx}
\usepackage[latin1]{inputenc}
\usepackage{xcolor,bm}
\usepackage{multirow,easyReview}
\usepackage[charter]{mathdesign}
\usepackage{amsmath}

\begin{document}

\hyphenation{a-long}

\title{Magnetic clusters in the paramagnetic phase of a high-temperature ferromagnetic\\ metal--organic framework}

\author{Giacomo~Prando}\email[E-mail: ]{giacomo.prando@unipv.it}\affiliation{Dipartimento di Fisica ``Alessandro Volta'', Universit\`a degli Studi di Pavia, 27100 Pavia, Italy}
\author{Benjamin~Costarella}\affiliation{Dipartimento di Fisica ``Alessandro Volta'', Universit\`a degli Studi di Pavia, 27100 Pavia, Italy}\affiliation{\'{E}cole Normale Sup\'{e}rieure Paris-Saclay, 91190 Gif-sur-Yvette, France}
\author{Matthew~S.~Dickson}\affiliation{Department of Chemistry, University of California, Berkeley, California 94720, United States}
\author{Ryan~A.~Murphy}\affiliation{Department of Chemistry, University of California, Berkeley, California 94720, United States}
\author{Jesse~G.~Park}\affiliation{Department of Chemistry, University of California, Berkeley, California 94720, United States}\affiliation{Department of Chemistry, Yonsei University, Seoul, 03722 Republic of Korea}
\author{Gianrico~Lamura}\affiliation{CNR-SPIN, 16152 Genova, Italy}
\author{Giuseppe~Allodi}\affiliation{Dipartimento di Scienze Matematiche, Fisiche ed Informatiche, Universit\`a di Parma, 43100 Parma, Italy}
\author{Cristian~Aloisi}\affiliation{Dipartimento di Fisica ``Alessandro Volta'', Universit\`a degli Studi di Pavia, 27100 Pavia, Italy}
\author{A\"eto~Apaix}\affiliation{Dipartimento di Fisica ``Alessandro Volta'', Universit\`a degli Studi di Pavia, 27100 Pavia, Italy}\affiliation{\'{E}cole Normale Sup\'{e}rieure de Lyon, 69342 Lyon, France}
\author{Maria~Cristina~Mozzati}\affiliation{Dipartimento di Fisica ``Alessandro Volta'', Universit\`a degli Studi di Pavia, 27100 Pavia, Italy}
\author{T.~David~Harris}\affiliation{Department of Chemistry, University of California, Berkeley, California 94720, United States}
\author{Jeffrey~R.~Long}\affiliation{Department of Chemistry, University of California, Berkeley, California 94720, United States}\affiliation{Department of Chemical and Biomolecular Engineering, University of California, Berkeley, California 94720, United States}\affiliation{Materials Sciences Division, Lawrence Berkeley National Laboratory, Berkeley, California 94720, United States}\affiliation{Department of Materials Science and Engineering, University of California, Berkeley, California 94720, United States}
\author{Pietro~Carretta}\affiliation{Dipartimento di Fisica ``Alessandro Volta'', Universit\`a degli Studi di Pavia, 27100 Pavia, Italy}

\begin{abstract}
	Owing to their exceptional chemical and electronic tunability, metal--organic frameworks can be designed to develop magnetic ground states making a range of applications feasible, from magnetic gas separation to the implementation of lightweight, rare-earth free permanent magnets. However, the typically weak exchange interactions mediated by the diamagnetic organic ligands result in ordering temperatures confined to the cryogenic limit. The itinerant magnetic ground state realized in the chromium-based framework Cr(tri)$_{2}$(CF$_{3}$SO$_{3}$)$_{0.33}$ (Htri, $1$\textit{H}-$1$,$2$,$3$-triazole) is a remarkable exception to this trend, showing a robust ferromagnetic behavior almost at ambient conditions. Here, we use dc SQUID magnetometry, nuclear magnetic resonance, and ferromagnetic resonance to study the magnetic state realized in this material. We highlight several thermally-activated relaxation mechanisms for the nuclear magnetization due to the tendency of electrons towards localization at low temperatures as well as the rotational dynamics of the charge-balancing triflate ions confined within the pores. Most interestingly, we report the development within the paramagnetic regime of mesoscopic magnetic correlated clusters whose slow dynamics in the MHz range are tracked by the nuclear moments, in agreement with the highly unconventional nature of the magnetic transition detected by dc SQUID magnetometry. We discuss the similarity between the clustered phase in the paramagnetic phase and the magnetoelectronic phase segregation leading to colossal magnetoresistance in manganites and cobaltites. These results demonstrate that high-temperature magnetic metal--organic frameworks can serve as a versatile platform for exploring correlated electron phenomena in low-density, chemically tunable materials.
\end{abstract}

\date{\today}

\maketitle

Metal--organic frameworks (MOFs) --- belonging to the more general class of coordination solids --- are composed of metal centers connected by organic polytopic ligands, resulting in materials with both chemical functionality and degrees of porosity that can be tuned depending on the vast choice of the structural elements \cite{Yag03,Fur13}. These properties make MOFs particularly attractive for applications ranging from gas storage and capture, gas sensing and separation, to catalysis. Saliently, they are also a convenient platform to realize amphidynamic crystals, characterized by both long-ranged structural order and the preservation of liquid-like molecular dynamics down to cryogenic temperatures \cite{Lie20,Per20a,Pra20,Don22}.

Because MOFs are highly chemically tunable, their electronic properties can be widely and predictably varied. In particular, MOFs have recently attracted attention as a possible platform to implement pre-determined magnetic ground states \cite{Tho20}. In addition to providing a platform for experiments of fundamental interest, this scenario offers exciting new avenues for applications. Indeed, combining precise control of the chemical environment of the metal centers --- which allows a careful tailoring of the magnetic anisotropy --- with the high structural porosity of MOFs could enable development of new lightweight, rare-earth-free permanent magnets \cite{Per20b} and facilitate magnetic gas separations \cite{Dec11,Tho20}. However, the realization of stable magnetic ground states with long-range order and high critical temperatures has been hampered by the typically weak magnetic coupling between metal centers separated by diamagnetic organic ligands \cite{Dec11,Tho20}.

\begin{figure}[t!]
	\centering
	\includegraphics[width=0.46\textwidth]{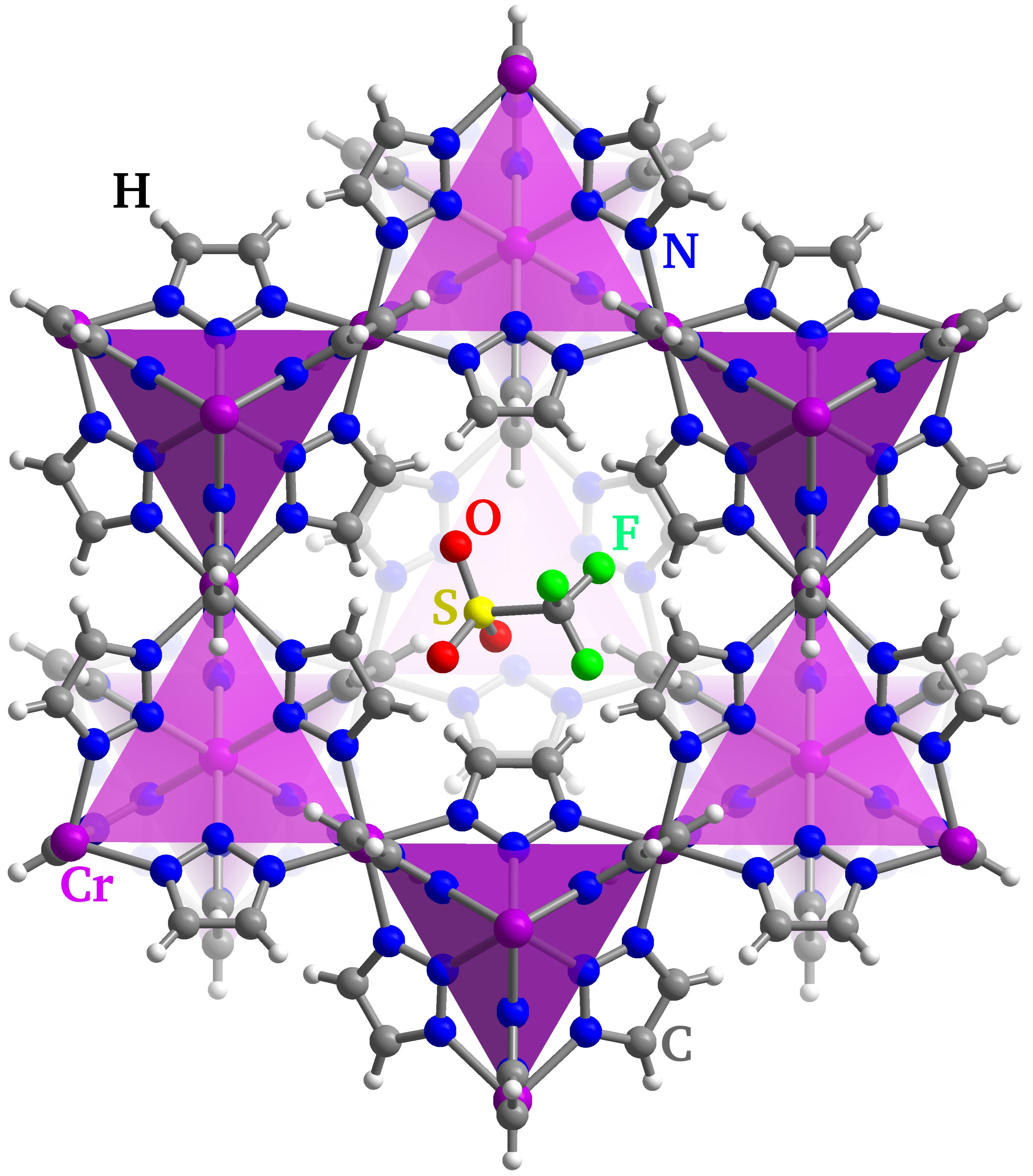}
	\caption{\label{FigStructure} Portion of the crystal structure of Cr(tri)$_{2}$(CF$_{3}$SO$_{3}$)$_{0.33}$ based on the high-temperature powder x-ray diffraction data reported in Ref.~\cite{Par21}. The image highlights a pyrochlore-like lattice made of triazolate-bridged chromium centers, as well as a disordered, charge-balancing trifluoromethanesulfonate anion in the framework pore. Cr, purple; S, yellow; F, green; O, red; N, blue; C, grey; H, white.}
\end{figure}
For this reason, the recent report of itinerant ferromagnetism with a high Curie temperature in the chromium-based MOF Cr(tri)$_{2}$(CF$_{3}$SO$_{3}$)$_{0.33}$ (Htri, $1$\textit{H}-$1$,$2$,$3$-triazole) is a particularly exciting development \cite{Par21}. In this material, Cr metal centers are octahedrally coordinated with $1$,$2$,$3$-triazolate ligands, leading to an overall pyrochlore-related structure \cite{Nut23} as depicted in Fig.~\ref{FigStructure}. The disordered triflate (CF$_{3}$SO$_{3}^{-}$) anion positioned at the center of the pore guarantees charge balance and, crucially, suggests that Cr ions are in a mixed-valence Cr$^{2+}$/Cr$^{3+}$ configuration \cite{Par21}. Mixed valency in the chromium centers favours ferromagnetic ordering via a double exchange mechanism, which makes the studied material the first example of a metal--organic framework exhibiting itinerant magnetism --- the most common source of magnetism in conventional solid state permanent magnets. Remarkably, this material displays a negative magnetoresistance of $\sim23\%$ at low temperatures, among the highest values for coordination solids \cite{Par21}.

In this work, we study the magnetic ground state realized in Cr(tri)$_{2}$(CF$_{3}$SO$_{3}$)$_{0.33}$ using dc SQUID magnetometry and from a local perspective using nuclear magnetic resonance on both {}$^{1}$H and {}$^{19}$F nuclei, complemented by ferromagnetic resonance spectroscopy. The scaling analysis of the isothermal dc magnetization curves highlights the highly unconventional nature of the magnetic transition. We find a complex dependence of the spin-lattice relaxation rate on temperature, which we interpret as the result of two independent thermally-activated relaxation mechanisms. In the first, electron hopping dynamically slows upon cooling, towards electron localization; in the second, the rotational dynamics of the charge-balancing triflate anion modulate relaxation. Remarkably, we find no evidence of critical dynamics at the magnetic transition. Instead, another Arrhenius-like process drives the nuclear relaxation well inside the paramagnetic state. We argue that these are evidence of the generation of mesoscopic magnetic clusters already within the paramagnetic state, akin to the magnetoelectronic phase segregation which has been interpreted as the origin of colossal magnetoresistance in manganese- and cobalt-based oxides. Although our dc SQUID magnetometry data rule out the development of a Griffiths-like state similar to what is realized in those complex oxides, we argue that the phenomenology of the clustered phase is still similar and can act as an intriguing analogy between inorganic oxides and magnetic metal--organic frameworks.

\section*{Results}\label{SectRes}

\subsection*{Static scaling of isothermal magnetization curves}

\begin{figure*}[t!]
	\centering
	\includegraphics[width=0.48\textwidth]{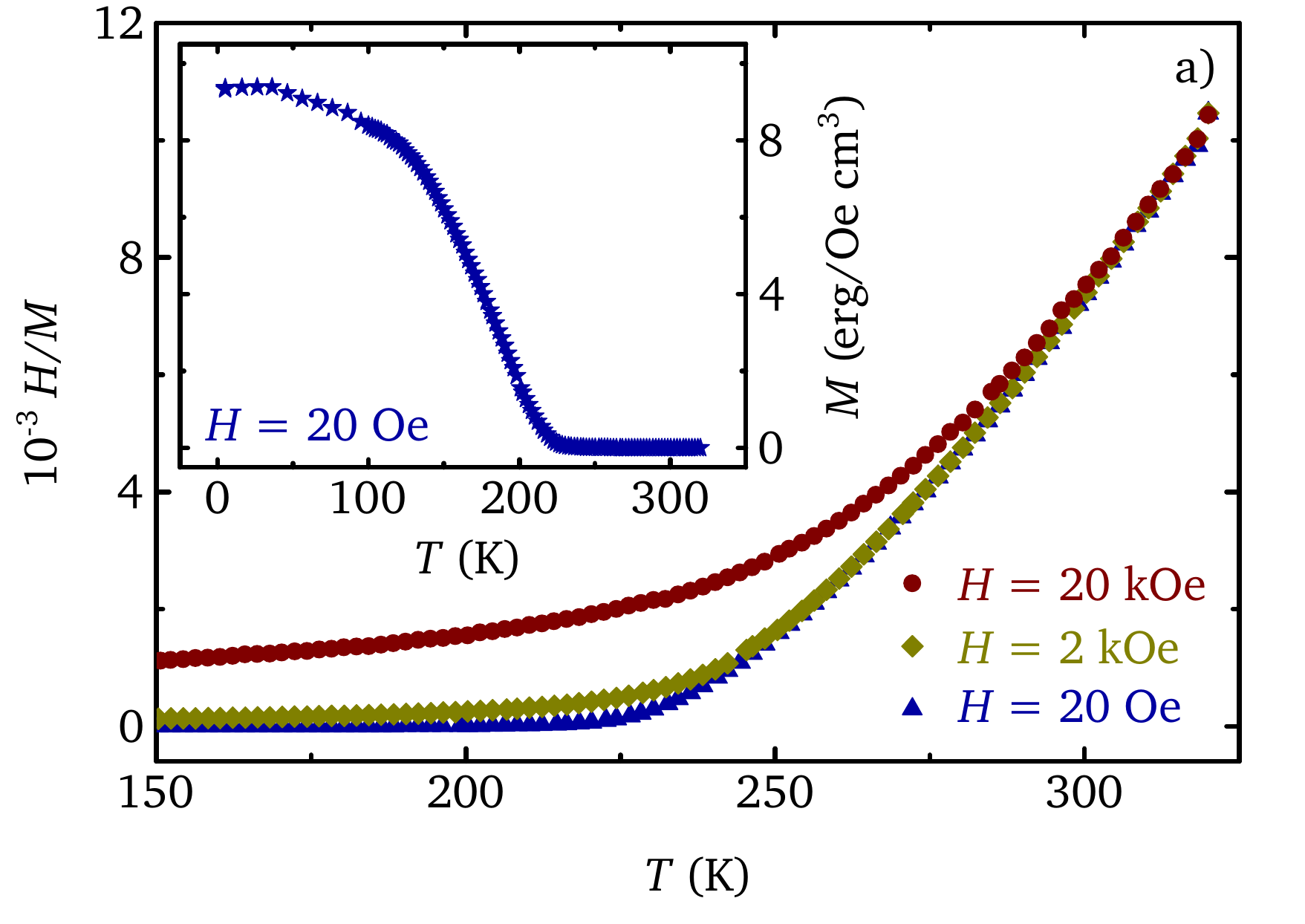} \hfill	\includegraphics[width=0.48\textwidth]{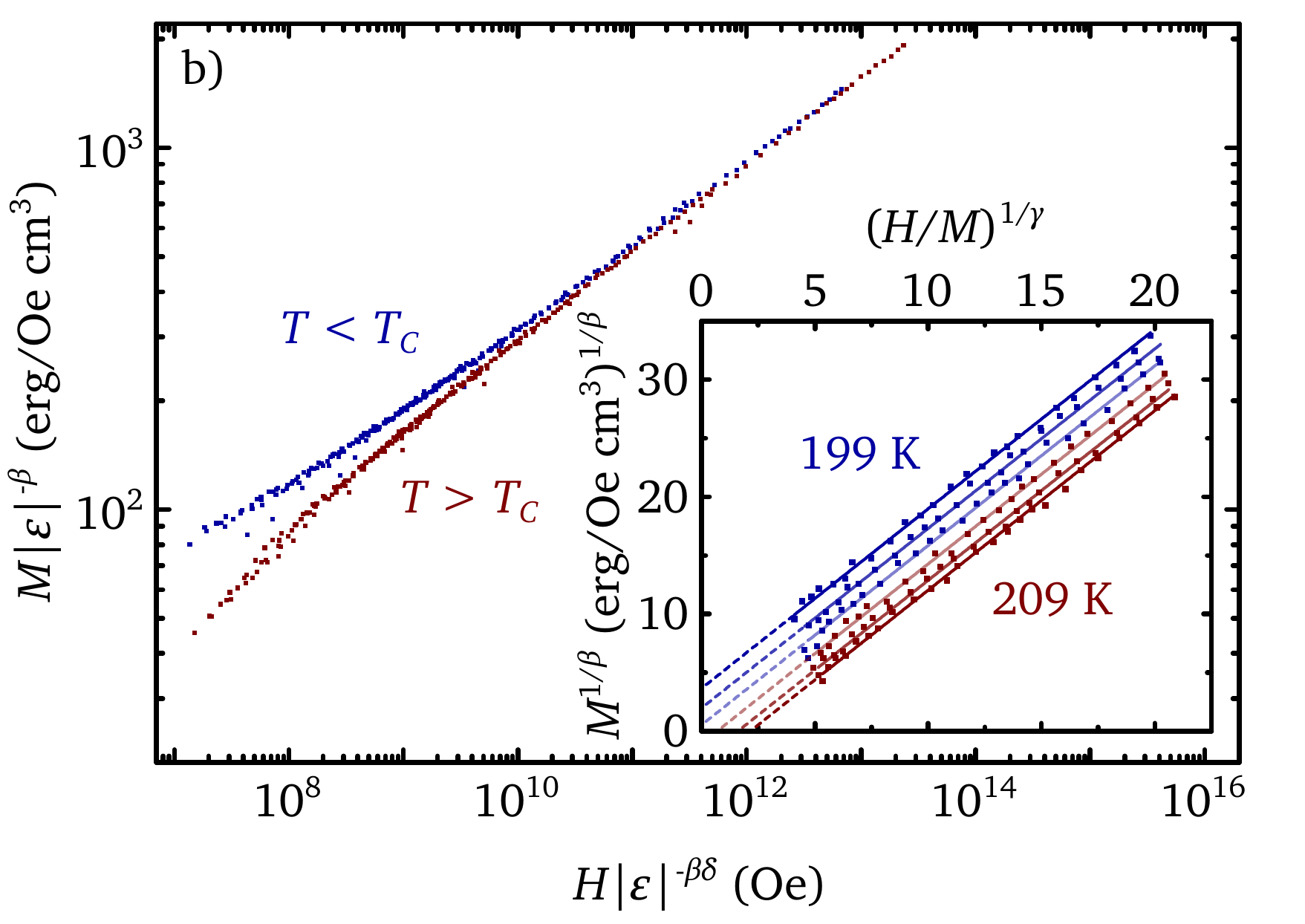}
	\caption{\label{FigMagnetization} a) Main panel: inverse susceptibility $H$/$M$ as a function of temperature after a zero-field-cooled protocol for three different values of fixed magnetic field. Inset: dependence of the magnetization $M$ on temperature at fixed magnetic field $H = 20$ Oe after a field-cooled protocol. b) Main panel: scaling analysis of the isothermal magnetization (see text). The $M$ vs $H$ curves measured at constant temperature values between $199$ K and $208.5$ K with $0.5$ K steps are reported. Inset: modified Arrott plot for the isothermal magnetization. The curves measured at constant temperature values between $199$ K and $209$ K are reported with $2$ K steps for the aim of clarity. The continuous lines are linear best-fits to the experimental data. The dashed lines are extrapolations of the linear fits.}
\end{figure*}
The dependence of the field-cooled dc magnetization on temperature (see Methods) for the investigated Cr(tri)$_{2}$(CF$_{3}$SO$_{3}$)$_{0.33}$ sample is reported in the inset to Fig.~\ref{FigMagnetization}a. The results display a close agreement with the data reported previously for a nominally identical sample \cite{Par21}, albeit with the abrupt rise in magnetization associated with the onset of ferromagnetism beginning at a somewhat lower temperature (we note that a modest batch-to-batch variation in ordering temperature is typical for this material). Additionally, as will be discussed later, the dependence of $H$/$M$ on temperature for different values of $H$ is smooth without any sharp drop in the paramagnetic regime. This is shown in the main panel of Fig.~\ref{FigMagnetization}a, where data measured under a zero-field-cooled protocol are reported. We stress that the results for field-cooled curves in this temperature region are completely equivalent to the zero-field-cooled curves.

The main panel of Fig.~\ref{FigMagnetization}b shows isothermal $M$ vs $H$ curves after a static-scaling analysis. Treating the relevant thermodynamic potential for the physical system under analysis as a generalized homogeneous function of temperature and magnetic field, it can be shown that in the critical region the following relation holds \cite{Sta71}
\begin{equation}\label{EqScaling}
	\frac{M}{\left|\varepsilon\right|^{\beta}} = F_{\pm}\left(\frac{H}{\left|\varepsilon\right|^{\beta\delta}}\right).
\end{equation}
Here, $\varepsilon$ is the reduced temperature
\begin{equation}
	\varepsilon = \frac{T - T_{C}}{T_{C}}
\end{equation}
while $\beta$ and $\delta$ are the critical exponents of the order parameter and of the critical isothermal $M$ vs $H$ curve, respectively. The intrinsic magnetic field $H$ already includes the correction for the demagnetizing factor for the considered sample geometry (see Methods). According to Eq.~\eqref{EqScaling}, after a proper normalization of both $M$ and $H$ based on the temperature value and on the critical exponents, the experimental isothermal curves should collapse to the empirical functions $F_{+}$ and $F_{-}$, corresponding to the conditions $T > T_{C}$ and $T < T_{C}$, respectively. Our results are reported in the main panel of Fig.~\ref{FigMagnetization}b and confirm the expected scaling using a value $T_{C} = 203.8$ K for the critical temperature and $\delta = 4.2$ and $\beta = 0.73$ for the two critical exponents.

An additional check on the values of the critical temperature and exponents estimated above comes from the use of the Arrott-Noakes approach \cite{Arr67}. In particular, when plotted on the so-called modified Arrott plot as $M^{1/\beta}$ vs $\left(H/M\right)^{1/\gamma}$, the isothermal curves at different temperatures should result in parallel straight lines with the critical isothermal curve passing through the axes' origin \cite{Tsu13,Tat14}. Here, the response function's critical exponent $\gamma = 2.336$ is obtained via the Widom relation $\beta \left(\delta - 1\right) = \gamma$ \cite{Sta71}. This analysis is shown in the inset to Fig.~\ref{FigMagnetization}b using the critical exponents estimated from the scaling analysis discussed above.

\subsection*{{}$^{1}$H and {}$^{19}$F nuclear magnetic resonance}

The crystallographic locations of the {}$^{1}$H and {}$^{19}$F nuclei probed by nuclear magnetic resonance (NMR) spectroscopy are illustrated in Fig.~\ref{FigStructure}. Here, hydrogen atoms are located on triazolate ligands within the MOF pyrochlore-like lattice, while the fluorine atoms are located in the charge-balancing triflate ions in the center of the pores. These nuclei, {}$^{1}$H and {}$^{19}$F, can thus serve as probes of the local magnetic fields in two distinct crystallographic environments.

\begin{figure}[t!]
	\centering
	\includegraphics[width=0.48\textwidth]{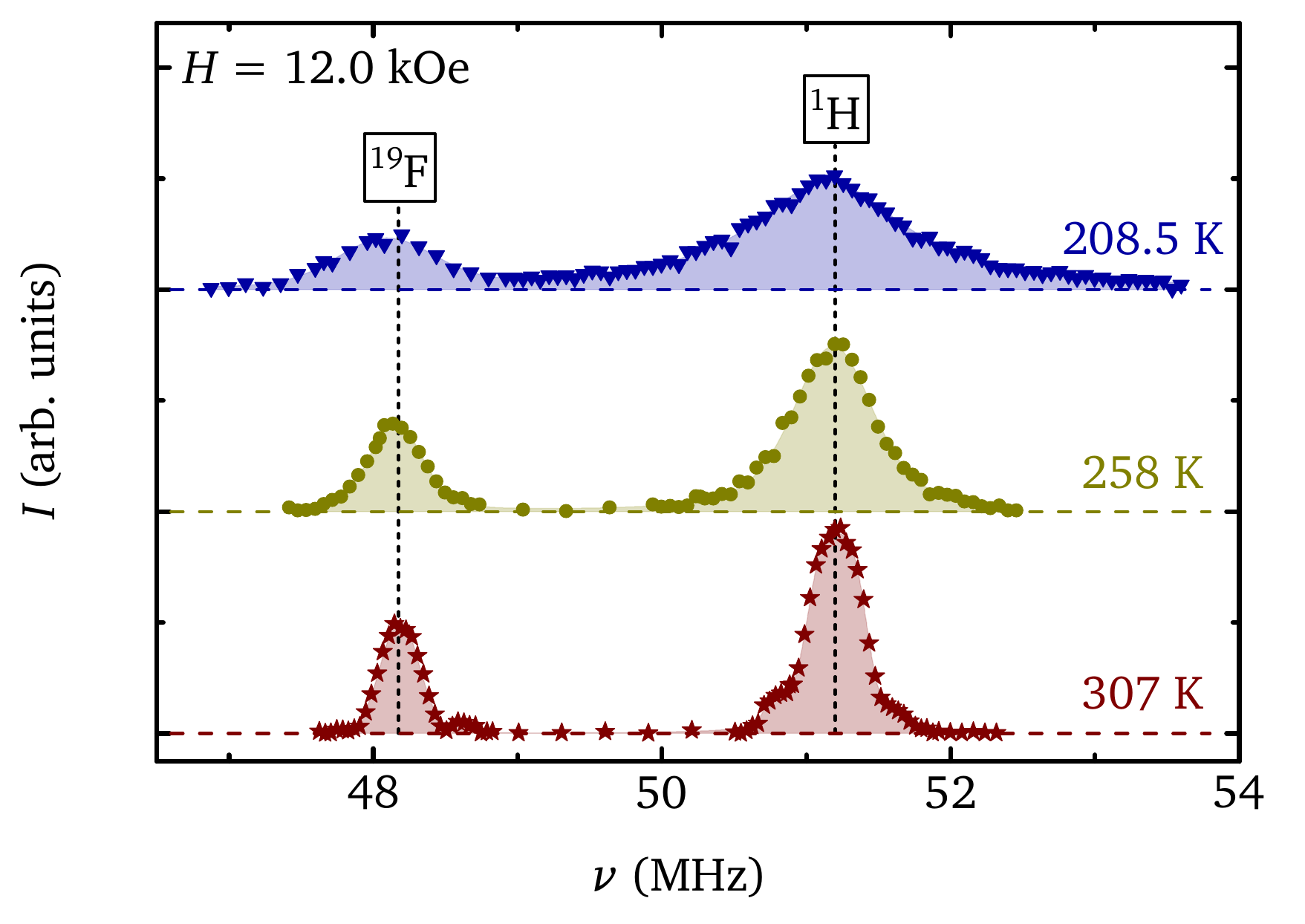}
	\caption{\label{FigSpectra} Representative frequency-swept, powder-averaged NMR spectra at a fixed magnetic field of $H = 12.0 \; \textrm{kOe}$ at selected temperatures in the paramagnetic regime. The different spectra are offset vertically for clarity. The two well-defined, inhomogeneously-broadened spectral lines are centered around the expected frequency values calculated from the gyromagnetic ratios characteristic of the {}$^{1}$H and {}$^{19}$F nuclei. The approach to the ferromagnetic phase ($T_{C} = 203.8$ K) while cooling induces an additional temperature-dependent line broadening which is much more marked than a minor paramagnetic line shift towards lower frequencies for both nuclei. The dashed vertical lines indicate the central resonance frequencies for {}$^{1}$H and {}$^{19}$F nuclei at $307$ K.}
\end{figure}

The NMR signal from {}$^{1}$H and {}$^{19}$F nuclei can be separately resolved in the measured frequency-swept spectra (Fig.~\ref{FigSpectra} --- see Methods). In particular, at fixed magnetic field, the spectra are characterized by two well-defined contributions which can be assigned unambiguously to the two different nuclei based on the characteristic values of the central frequency. Notably, both signals are progressively broadened on approaching the ferromagnetic ordering temperature from the paramagnetic regime. We interpret this marked dependence of the inhomogeneous, powder-averaged spectral broadening as clear evidence of the development of internal magnetic fields associated with the ferromagnetic state at the sites of both {}$^{1}$H and {}$^{19}$F nuclei. A paramagnetic shift of the spectral lines towards lower frequencies can be distinguished upon decreasing temperature, though the effect is much less marked than the spectral broadening.

\begin{figure}[t!]
	\centering
	\includegraphics[width=0.48\textwidth]{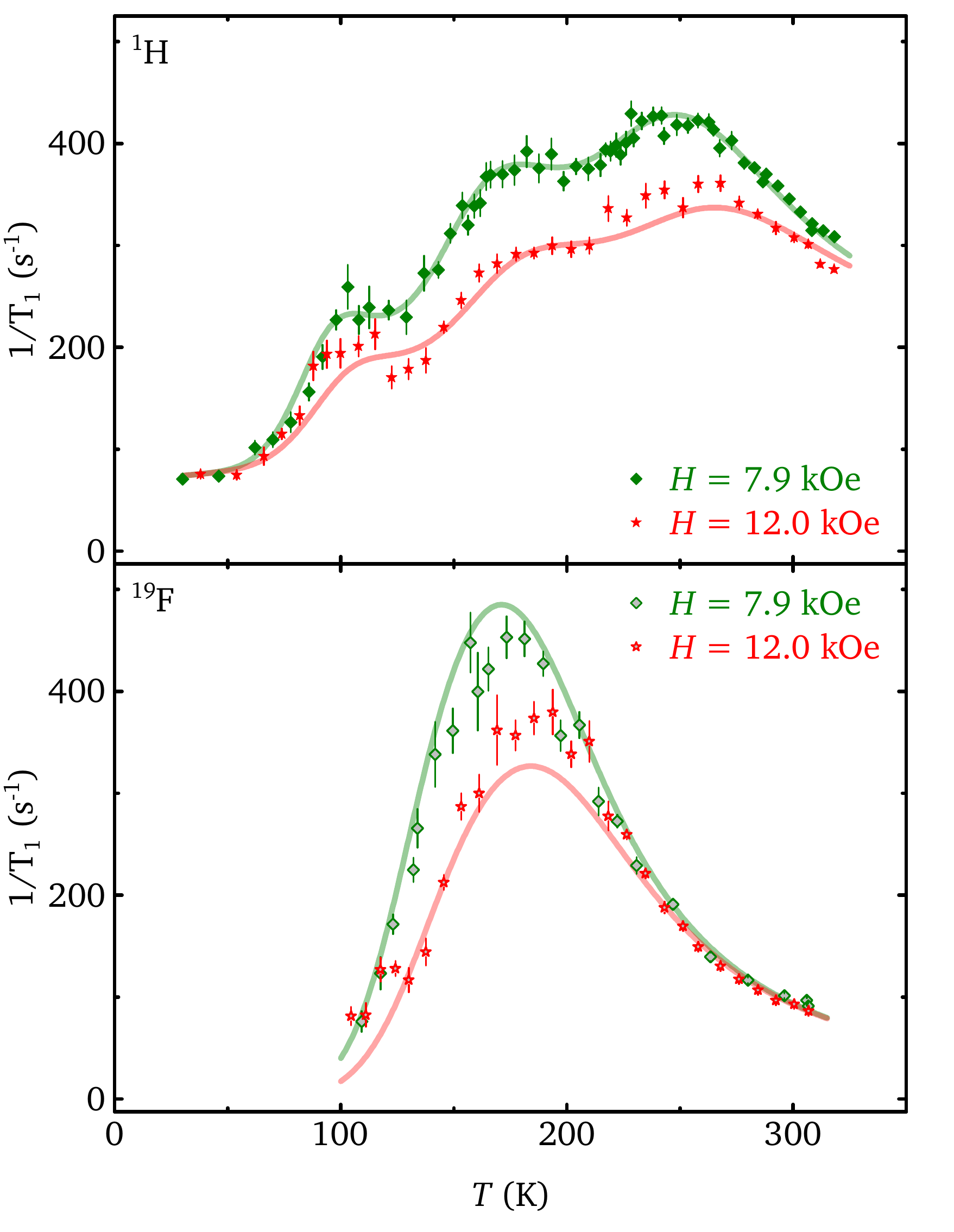}
	\caption{\label{Fig1overT1vsT} Dependence of the spin-lattice relaxation rate for the nuclear magnetization of {}$^{1}$H and {}$^{19}$F nuclei (upper and lower panels, respectively) on temperature at two different external magnetic fields. The error bars are the standard deviation of the fit parameters in Eq.~\eqref{EqRecoveryFit} (see Methods). The continuous lines are the results of global fittings based on Eq.~\eqref{EqDistrBPP} (lower panel) and Eq.~\eqref{EqThreeBPP} (upper panel) using the Larmor frequencies at the different magnetic fields as fixed parameters and the remaining quantities as shared parameters (see text).}
\end{figure}

We quantified the spin-lattice relaxation time, $\textrm{T}_{1}$, for the nuclear magnetization of both {}$^{1}$H and {}$^{19}$F nuclei (see Methods) as a function of temperature and at different values of the external magnetic field. The results for $H = 7.9 \; \textrm{kOe}$ and $H = 12.0 \; \textrm{kOe}$ are summarized in Fig.~\ref{Fig1overT1vsT}. For both nuclei and at all temperature values, $\textrm{T}_{1}$ is of the order of a few ms, which is indicative of intense fluctuating magnetic fields at both nuclear sites. Remarkably, for both nuclei, we do not observe any sharp critical peak in the spin-lattice relaxation rate $1/\textrm{T}_{1}$, as is typically found for ferromagnetic phase transitions. The experimental results for the {}$^{19}$F nuclei instead show a broad peak for $1/\textrm{T}_{1}$ centered around $170 - 190$ K, well below the critical temperature for the magnetic phase transition. The dependence of $1/\textrm{T}_{1}$ on temperature for the {}$^{1}$H nuclei is more complicated. Here, a local maximum is observed at $\sim110$ K, followed by two more pronounced broad peaks at $170 - 190$ K and $230 - 250$ K. Comparing data taken at $H = 7.9 \; \textrm{kOe}$ and $H = 12.0 \; \textrm{kOe}$, the larger magnetic field shifts the position of all the observed maxima to higher temperature while also suppressing the absolute value of $1/\textrm{T}_{1}$ for both {}$^{1}$H and {}$^{19}$F.

Based on this qualitative behavior of the results, we attempted a best-fit of the experimental data using the Bloembergen, Purcell, and Pound (BPP) model for the {}$^{19}$F nuclear spin-lattice relaxation \cite{Blo48,Abr61,Sli90} (see Methods). In particular, the expression
\begin{align}\label{EqDistrBPP}
	\textrm{T}_{1}^{-1}\left(T\right) =& \frac{C \; T}{4 \omega_{L} \delta\vartheta} \left\{\arctan\left[\sinh\left(\frac{\vartheta+\delta\vartheta}{T}+\ln(\omega_{L}\tau_{0})\right)\right]\right.\nonumber\\ &-\left. \arctan\left[\sinh\left(\frac{\vartheta-\delta\vartheta}{T}+\ln(\omega_{L}\tau_{0})\right)\right]
	\right\}
\end{align}
was used for fitting, also accounting for a flat distribution of the activation barrier between $\vartheta - \delta\vartheta$ and $\vartheta + \delta\vartheta$ (see Ref.~\cite{Pra23}). The curves reported in the lower panel of Fig.~\ref{Fig1overT1vsT} are the results of a global best-fitting procedure, where the experimental Larmor angular frequencies $\omega_{L}$ are fixed parameters for the two different datasets while the other quantities ($C$, $\tau_{0}$, $\vartheta$, and $\delta\vartheta$) are shared parameters varied simultaneously for the two datasets during the numerical least-square minimization. The resulting values of the parameters are reported in Tab.~\ref{TabFitF}.
\begin{figure*}[t!]
	\centering
	\includegraphics[width=0.48\textwidth]{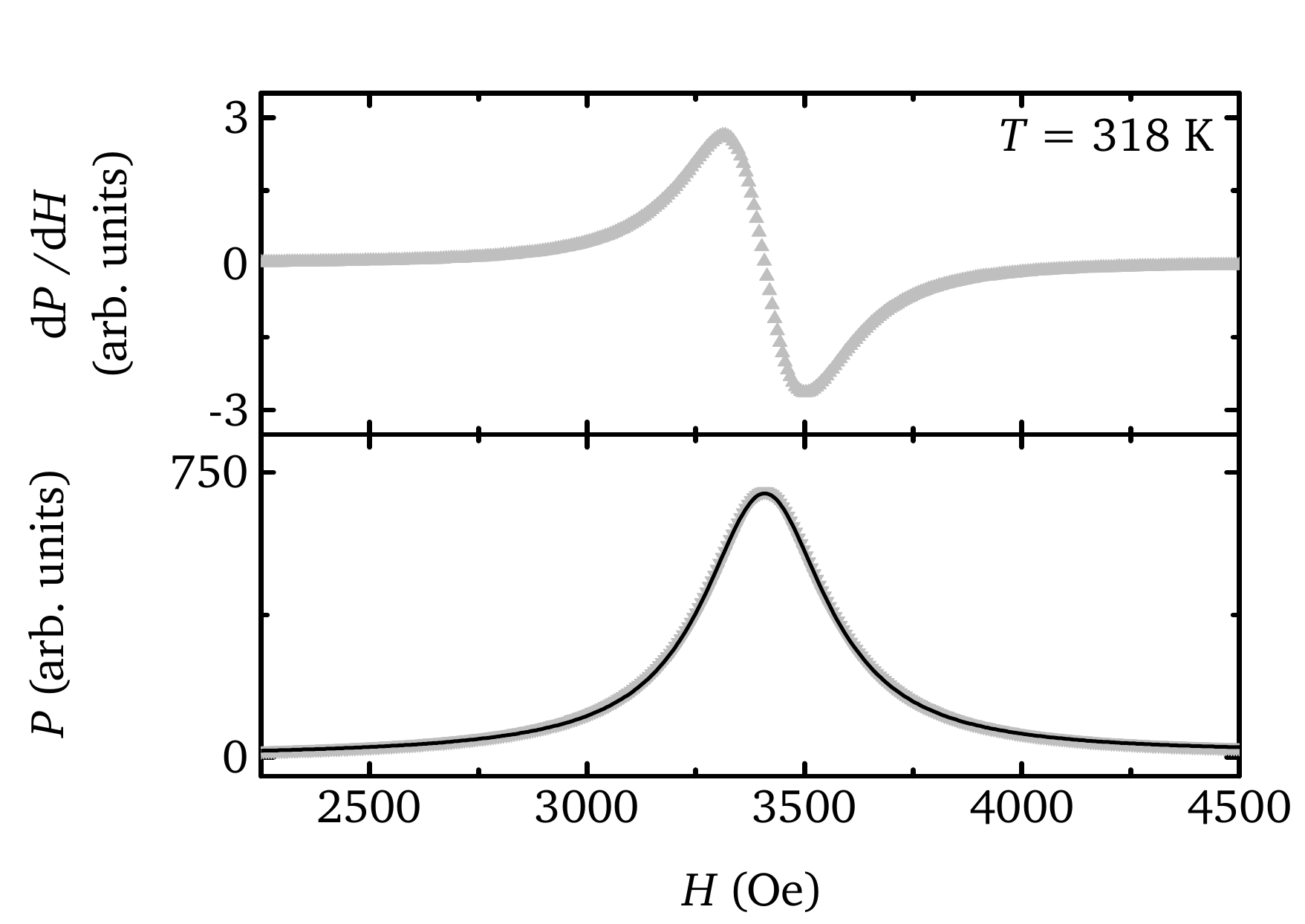} \hfill \includegraphics[width=0.48\textwidth]{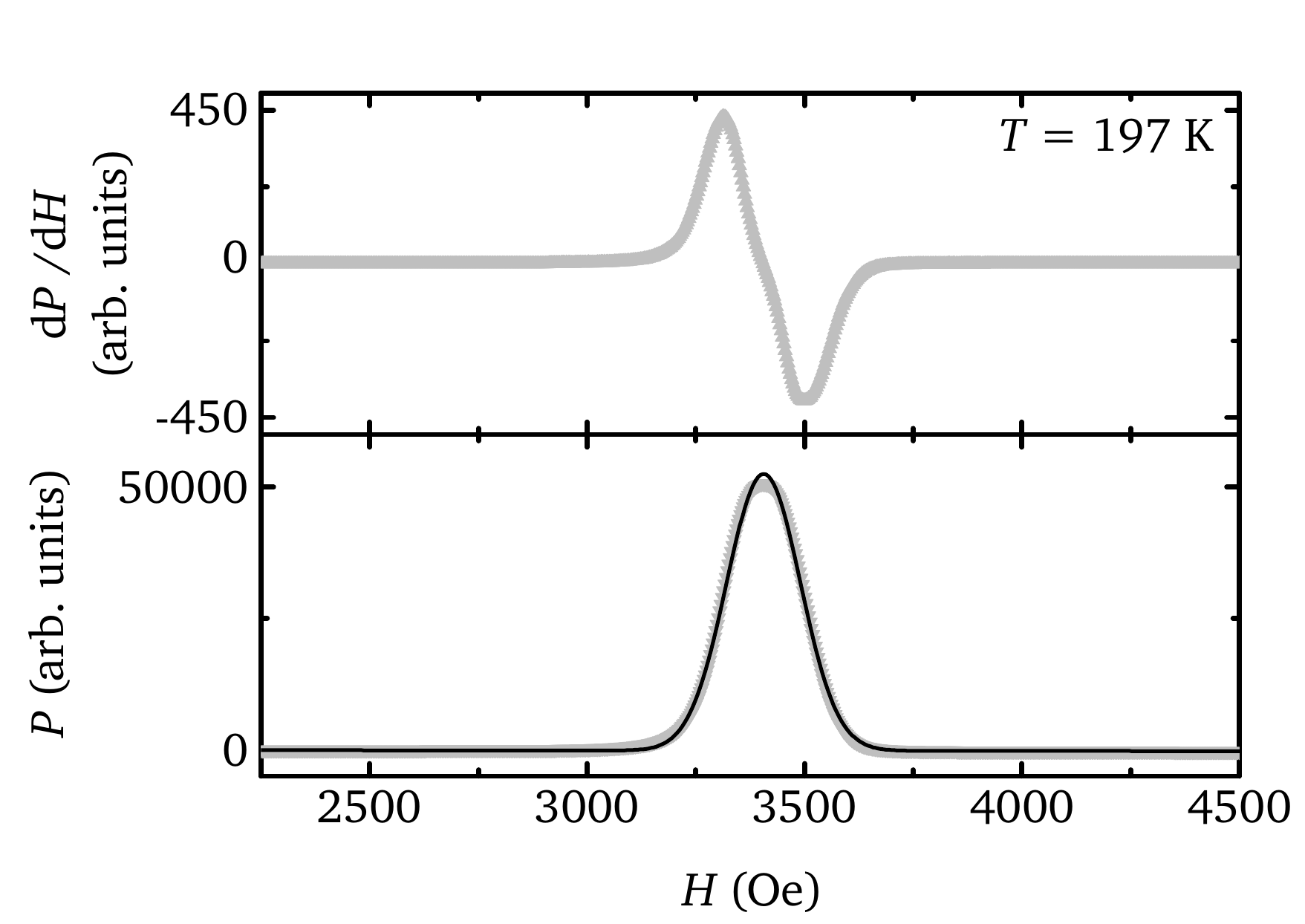}
	\caption{\label{FigFMRspectra} X-band FMR spectra at the representative temperatures of $T = 318$ K and $T = 197$ K (left-hand panel and right-hand panel, respectively). For both temperatures, the first-derivative experimental data and their numerical integration are reported in the upper and lower panels, respectively. The continuous lines in the lower panels are based upon a best-fitting function according to Eq.~\eqref{EqVoigt}. The markedly exchange-narrowed, Cauchy-Lorentz character of the high-temperature spectrum is turned to a Gaussian-like shape at low temperatures.}
\end{figure*}

\begin{table}[b!]
	\begin{tabular}{c  c  c  c  c  c  c}
		\hline
		\hline
		$C$ ($10^{11}$ s$^{-2}$) & \phantom{aa} & $\tau_{0}$ ($10^{-11}$ s) & \phantom{aa} & $\vartheta$ (K) & \phantom{aa} & $\delta\vartheta$ (K)\\
		\hline
		$2.35 \pm 0.05$ & \phantom{aa} & $1.4 \pm 0.1$ & \phantom{aa} & $990 \pm 15$ & \phantom{aa} & $205 \pm 15$\\
		\hline
		\hline
	\end{tabular}
	\caption{\label{TabFitF} Results of the global best-fitting procedure based on Eq.~\eqref{EqDistrBPP} of the $1/\textrm{T}_{1}$ data for the {}$^{19}$F nuclear magnetization shown in the lower panel of Fig.~\ref{Fig1overT1vsT}.}
	\vspace*{0.3cm}
	\begin{tabular}{c  c  c  c  c  c  c }
		\hline
		\hline
		{} & \phantom{aa} & $C$ ($10^{10}$ s$^{-2}$) & \phantom{aa} & $\tau_{0}$ ($10^{-11}$ s) & \phantom{aa} & $\vartheta$ (K)\\
		\hline
		LT & \phantom{aa} & $5.65 \pm 0.25$ & \phantom{aa} & $3.4 \pm 1.7$ & \phantom{aa} & $500 \pm 50$\\
		\hline
		IT & \phantom{aa} & $9.0 \pm 0.5$ & \phantom{aa} & $1.8 \pm 0.6$ & \phantom{aa} & $960 \pm 60$\\
		\hline
		HT & \phantom{aa} & $9.4 \pm 0.4$ & \phantom{aa} & $1.20 \pm 0.15$ & \phantom{aa} & $1525 \pm 35$\\
		\hline
		\hline
	\end{tabular}
	\caption{\label{TabFitH} Results of the global best-fitting procedure based on Eq.~\eqref{EqThreeBPP} for the $1/\textrm{T}_{1}$ data for the {}$^{1}$H nuclear magnetization shown in the upper panel of Fig.~\ref{Fig1overT1vsT}. Also, the values $K = 0.210 \pm 0.025 \; \left(\textrm{K s}\right)^{-1}$ and $b = 68.0 \pm 2.5 \; \textrm{s}^{-1}$ are estimated from the global best-fitting procedure.}
\end{table}
For the spin-lattice relaxation rate of the {}$^{1}$H nuclear magnetization, a more elaborate fitting function was necessary. In particular, we treat the spin-lattice relaxation rate as the sum of three independent BPP contributions at low-, intermediate-, and high-temperature (LT, IT, and HT, respectively) as follows
\begin{align}\label{EqThreeBPP}
	\textrm{T}_{1}^{-1}\left(T\right) = & \; \textrm{T}_{1}^{-1}\left(T\right)_{\rm{LT}}^{\rm{BPP}} + \textrm{T}_{1}^{-1}\left(T\right)_{\rm{IT}}^{\rm{BPP}} + \textrm{T}_{1}^{-1}\left(T\right)_{\rm{HT}}^{\rm{BPP}}\nonumber\\ &+ KT + b
\end{align}
where a linear-in-temperature, Korringa-like contribution and a constant term are added. The best-fits are shown in the upper panel of Fig.~\ref{Fig1overT1vsT} and are in satisfactory agreement with the experimental data. In order to limit the number of free parameters in the fitting procedure, we refrained from introducing any flat distribution of activation energies. At the same time, we used again a global best-fitting procedure, sharing all the parameters (except the fixed Larmor angular frequencies) between the two datasets during the numerical least-square minimization. The resulting values of the best-fitting parameters are reported in Tab.~\ref{TabFitH}.

\subsection*{Ferromagnetic resonance}

Additional information on the properties of the ferromagnetic state in the MOF was obtained by means of continuous-wave, X-band ferromagnetic resonance (FMR) experiments (see Methods). Representative FMR spectra at different temperatures are reported in Fig.~\ref{FigFMRspectra}, evidencing a single, broad spectral contribution around the expected resonant magnetic field for the free electron at the working frequency. Although the central field is not dependent on temperature within the experimental uncertainty, the spectral width does depend strongly on temperature. The absolute values at the minimum and maximum of the first-derivative d$P$/d$H$ vs $H$ curves almost coincide and, accordingly, the spectral line shows no Dyson-like distortion. Dysonian distortion arises from an admixture of the absorptive and dispersive components and is typically observed in metallic samples, implying that in the studied MOF the skin depth is thicker than the typical grain size \cite{Bar81,Pra16b}. The numerical integration of the first-derivative d$P$/d$H$ vs $H$ data evidences a clear crossover from an exchange-narrowed, Cauchy-Lorentz-like regime to a Gaussian-like regime for the spectral shape at high and low temperatures, respectively. The crossover is gradual on decreasing the temperature from $\sim270$ K to $\sim220$ K and in this window the spectrum is properly described as an admixture of Cauchy-Lorentz and Gaussian shapes.

Based on these observations, we used the Voigt function
\begin{align}\label{EqVoigt}
	y =& \; \frac{2 \; A}{\Gamma_{G}} \; \left(\frac{\ln(2)}{\pi}\right)^{1/2} \frac{a}{\pi} \; \int_{-\infty}^{+\infty} \frac{\exp\left(-t^{2}\right)}{a^{2} + \left(v- t\right)^{2}} \; \textrm{d}t\nonumber\\& + B \cdot H + y_{0}
\end{align}
to fit the numerically-integrated experimental data. Here, the parameters $B$ and $y_{0}$ account for a linear background while
\begin{equation}
	a = \sqrt{\ln(2)} \; \frac{\Gamma_{L}}{\Gamma_{G}} \qquad \textrm{and} \qquad v = \frac{2 \sqrt{\ln(2)}}{\Gamma_{G}} \left(H - H_{c}\right).
\end{equation}
The Voigt function is centered at $H_{c}$ and it convolutes a Cauchy-Lorentz function with a Gaussian function with full widths at half maximum (FWHM) $\Gamma_{L}$ and $\Gamma_{G}$, respectively. Based on this fitting function, we extract the relevant parameters for the Voigt function, \textit{i.e.}, the subtended area $A$ and the FWHM approximated as \cite{Whi68,Kie73,Oli77}
\begin{equation}\label{EqVoigtFWHM}
	\Gamma_{V} \simeq 0.5346 \; \Gamma_{L} + \sqrt{0.2166 \; \Gamma_{L}^{2} + \Gamma_{G}^{2}}.
\end{equation}

\begin{figure}[t!]
	\centering
	\includegraphics[width=0.48\textwidth]{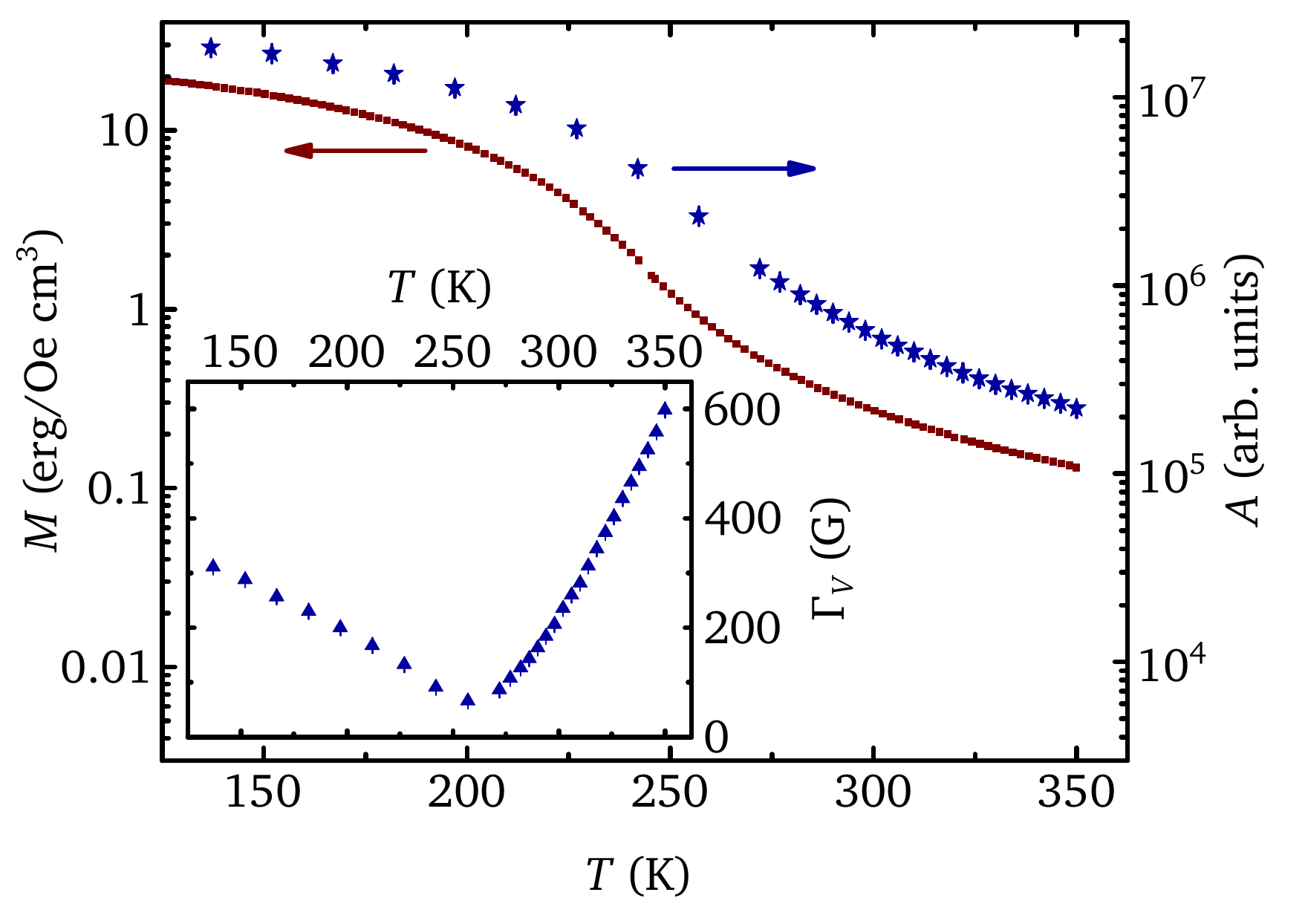}
	\caption{\label{FigFMRamplitude} Main panel: dependence of the area subtended by the numerically-integrated experimental data on temperature (blue stars, values on the right-hand axis). The red squares represent the dependence of the dc magnetization as a function of temperature at a fixed magnetic field of $H = 2$ kOe (values on the left-hand axis). Inset: dependence of the FWHM of the Voigtian fitting curves based on Eq.~\eqref{EqVoigtFWHM} (see text). In both panels, the error bars are the standard deviation of the fit parameters in Eq.~\eqref{EqVoigt}.}
\end{figure}

The results of the best-fitting procedure based on Eq.~\eqref{EqVoigt} are generally very good, in spite of minor discrepancies in the low-temperature regime around the center of the spectrum (see Fig.~\ref{FigFMRspectra}). The dependence of the fitting parameters on temperature is reported in Fig.~\ref{FigFMRamplitude}. The evolution of the area $A$ on varying temperature mimics precisely that of the dc magnetization at comparable magnetic field, as expected in FMR. At the same time, the dependence of the FWHM also resembles what is routinely measured for FMR across the transition from a paramagnetic to a magnetically-ordered state. In particular, the linewidth decreases steadily on decreasing temperature in the paramagnetic regime until a minimum is reached at temperatures well above the critical one \cite{Tay75,Kot78,Hub99,Ehl12,Pra16b}, even though a critical divergence is not observed \cite{Kot78,Ehl12}. Eventually, on further decreasing the temperature, the spectra become broader. We note that no anomaly is observed at around $170 - 190$ K, where both {}$^{1}$H and {}$^{19}$F NMR spectroscopy detect a major contribution to the spin-lattice relaxation rate.

\section*{Discussion}\label{SectDisc}

\subsection*{Charge localization}

We discuss our interpretation of the experimental results starting from the bump observed in the {}$^{1}$H spin-lattice relaxation rate at around $110$ K. This anomaly is located a few tens of Kelvin above the observation of a sudden drop in the electrical conductivity of the material \cite{Par21}. Accordingly, we argue that the spin-lattice relaxation in this temperature regime is favoured dynamic slowing of the charge hopping process, leading to localization. In particular, when the characteristic time of this hopping process matches the inverse of the Larmor frequency, the induced time-varying local magnetic fields can be effective in driving the relaxation of the nuclear magnetization. A similar mechanism was discussed in itinerant ferromagnetic manganites \cite{Sav99}. Here, it was suggested that the decrease of the correlation time upon decreasing temperature for the considered manganite was clear evidence that the hopping charges could be described as spin polarons. Small polaron formation was indeed previously invoked to explain the electronic conductivity of Cr(tri)$_{2}$(CF$_{3}$SO$_{3}$)$_{0.33}$; however, the more conventional increase of $\tau_{c}$ upon decreasing temperature observed here does not alone allow us to reach the same conclusion of spin polaron formation \cite{Par21}.

\subsection*{Molecular dynamics of the triflate anion}

We interpret the marked maxima observed in the spin-lattice relaxation rate of both {}$^{1}$H and {}$^{19}$F nuclei at around $170 - 190$ K in terms of the molecular motion of the charge-balancing triflate anion and, in particular, of the rotational motion of the CF$_{3}$ group about its $C_{3}$ axis. We start from the observation that, in this temperature regime, both {}$^{1}$H and {}$^{19}$F nuclei exhibit similar dynamics, as is evident based on a comparison of the estimated $\tau_{0}$ and $\vartheta$ values in Tab.~\ref{TabFitF} and Tab.~\ref{TabFitH}. These dynamics do not have a magnetic origin, as there is no anomaly in either the dc magnetization or the amplitude/linewidth of the FMR signal in this temperature range.

{}$^{19}$F NMR measurements on diamagnetic polymers containing triflate moieties help guide our interpretation. In this literature, a characteristic minimum in T$_{1}$ is observed in the same temperature range as in our measurements, with reported values for $\tau_{0}$ and $\vartheta$ in good quantitative agreement with our findings \cite{Miz90,Iid92}. At the same time, T$_{1}$ values for these diamagnetic polymers are longer by two orders of magnitude. We explain this discrepancy by noting that in Cr(tri)$_{2}$(CF$_{3}$SO$_{3}$)$_{0.33}$ the rotational motion of the CF$_{3}$ group about its $C_{3}$ axis takes place in a ferromagnetic background and, accordingly, the probed fluctuating local magnetic fields are more intense. Interestingly, while rotational dynamics would in principle primarily affect {}$^{19}$F nuclei, inhomogeneous spectral broadening leads the {}$^{19}$F and {}$^{1}$H signals to partly superimpose already at around $200$ K, as shown in Fig.~\ref{FigSpectra}. This makes spin-diffusion processes between the two nuclear species possible, leading to an intense effect on the spin-lattice relaxation rate of the {}$^{1}$H nuclear magnetization as well. The emergence of molecular dynamics setting in below the magnetic ordering temperature is unique, to the best of our knowledge. These findings highlight the utility of MOFs as platforms for studying unusual combinations of phenomena such as molecular dynamics inside of a magnet.

\subsection*{Preformed magnetic clusters in the paramagnetic regime}

As already noted, the temperature dependence of the spin-lattice relaxation rate does not show clear evidence of critical slowing down of fluctuations, whose signature would be a sharp divergence of $1/$T$_{1}$ at the critical temperature. The same argument holds for the FMR linewidth. These would be conventional results for a ferromagnetic phase transition, in contrast with the unexpected activated dynamics probed by {}$^{1}$H nuclei at around $230 - 250$ K. We stress that in this high-temperature regime the system does not display bulk magnetic order, as confirmed by the dc SQUID magnetometry data. Accordingly, the amplitude of the local modulated magnetic field would arise from the nuclear dipole-dipole interaction and, in this case, the T$_{1}$ values would be at least a few orders of magnitude longer than what we find experimentally \cite{Per20a}.

Additional clues to the origins of the slow thermally-activated dynamics in the paramagnetic state can be obtained from the values of the critical exponents estimated from the scaling analysis discussed above. In particular, the estimated values are quite unusual. Within the $3$D Heisenberg universality class the values $0.36$ and $4.86$ would be expected for $\beta$ and $\delta$, respectively --- our experimental value of $\beta = 0.73$ is far from expectation. For comparison, mean-field theories predict $\beta_{\rm{MF}} = 0.5$. Unusually high values for $\beta$ have been observed in magnetic systems in the presence of preformed mesoscopic magnetic clusters in the paramagnetic regime \cite{Sah22,Cha23}. This suggests that the same phenomenology is at work in the currently studied MOF, where the {}$^{1}$H spin-lattice relaxation rate in the high temperature regime would then be dominated by the slow dynamics of the mesoscopic clusters.

This kind of magnetoelectronic phase segregation above the critical temperature is ubiquitous in transition-metal oxides such as manganites and cobaltites \cite{Mor99,Ueh99,Dag01,Dag05}, marking yet another interesting analogy of these oxides with Cr(tri)$_{2}$(CF$_{3}$SO$_{3}$)$_{0.33}$. It should be remarked that the clustered state in oxides is often described in terms of the Griffiths model, originally formulated for the progressive magnetic dilution of a ferromagnetic ensemble of Ising moments \cite{Gri69}, which has been interpreted as the very origin of the colossal magnetoresistance in oxides \cite{Sal02}. One of the fingerprints of the Griffiths phase is the sudden drop of the low-field magnetic susceptibility well above the critical temperature \cite{Sal02,Shi06,Hua06}, a feature that is not observed in our experimental data shown in Fig.~\ref{FigMagnetization}b. However, the absence of a Griffiths state does not imply the absence of a clustered state, as non-Griffiths clustered phases have been reported in oxides as well \cite{He07}. In any case, the rich magnetoelectronic physics of Cr(tri)$_{2}$(CF$_{3}$SO$_{3}$)$_{0.33}$ demonstrates that molecule-based materials can host phenomena typically associated with inorganic solid state materials.

\section*{Conclusions}

We report a detailed study of the magnetic ground state in a Cr-based metal--organic framework using dc SQUID magnetometry as well as nuclear magnetic resonance and ferromagnetic resonance. The dependence of the spin-lattice relaxation on temperature for {}$^{1}$H nuclei --- located on the ligands within the metal--organic framework lattice --- is complex and can be described in terms of three independent thermally-activated processes. One of these is also reflected in the spin-lattice relaxation of the magnetization of {}$^{19}$F nuclei, located in the charge-balancing triflate anion in the center of the pores, and is associated with the rotational molecular dynamics of the anion itself. Associating the two other processes to the slowing down of the electron hopping at low temperatures and to the emergence of correlated magnetic clusters within the paramagnetic state stresses a strong qualitative analogy of the studied material with complex manganese- and cobalt-based oxides. The results warrant further investigations with particular reference to the high magnetoresistance shown by this Cr-based metal--organic framework.

\section*{Methods}

\subsection*{Synthesis}

All manipulations were performed under a dry Ar atmosphere. The compound Cr(CF$_{3}$SO$_{3}$)$_{2}$ was prepared according to a literature procedure \cite{Dix90}. Anhydrous N,N-dimethylformamide (DMF) and dichloromethane were dried and degassed using a JC Meyer solvent system and stored over $3$-{\AA} molecular sieves prior to use. Liquid $1H$-$1$,$2$,$3$-triazole ($\geq 99\%$ purity, AFG Bioscience) was degassed via four freeze-pump-thaw cycles and stored over $3$-{\AA} molecular sieves. The compound Cr(tri)$_{2}$(CF$_{3}$SO$_{3}$)$_{0.33}$ was synthesized according to a literature procedure \cite{Par21} with modifications. A blue $0.5$ M solution of Cr(CF$_{3}$SO$_{3}$)$_{2}$ ($28.00$ g, $79.98$ mmol) in DMF ($160$ mL) was added to $1H$-$1$,$2$,$3$-triazole ($14.0$ mL, $242$ mmol) and divided into sixteen $20$ mL scintillation vials. The vials were sealed with polytetrafluoroethylene-lined caps and heated at $120~^{\circ}\textrm{C}$ for $4$ days. The resulting dark purple suspensions were then cooled to room temperature and combined. Given the tendency of the fine suspension to penetrate polymer filter membranes, purification was achieved by a washing procedure that made use of the magnetic properties of the material. A portion of solvent ($2 \times 100$ mL DMF, $100$ mL dichloromethane, $4 \times 50$ mL dichloromethane) was added to the solid via cannula transfer on a Schlenk line under Ar, and after each addition the suspension was cooled to near the magnetic ordering temperature of Cr(tri)$_{2}$(CF$_{3}$SO$_{3}$)$_{0.33}$ ($\sim220$ K) using a cooling bath of dry ice in equal parts ethanol and ethylene glycol. A powerful $2.5$-cm neodymium magnet was placed against the outer wall of the sample flask to attract the solid, allowing removal of only the liquid via cannula transfer. After the washings were complete, the product was dried under dynamic vacuum ($< 20 \; \mu$bar) at $130~^{\circ}\textrm{C}$ for $18$ h to afford $2.664$ g ($14\%$) of dark purple microcrystalline powder. Elemental analyses for C, H, and N were obtained from the Microanalytical Laboratory at the University of California, Berkeley. Calculated for CrC$_{4.33}$H$_{4}$N$_{6}$FS$_{0.33}$O: C, $21.91$; H, $1.70$; N, $35.41$. Found: C, $22.29$; H, $1.84$; N, $34.00$.

\subsection*{Dc magnetometry}

We performed dc magnetometry measurements using a MPMS2 SQUID magnetometer (Quantum Design) on unrestrained powders of the investigated MOF sealed in a cylindrical quartz ampoule under an inert He atmosphere. The applied magnetic field was parallel to the symmetry axis of the ampoule. We measured the magnetic moment of the sample as a function of temperature under both zero-field-cooled and field-cooled protocols, while every isothermal curve was recorded after a zero-field-cooled protocol from $320$ K to the target temperature.

In order to calculate the demagnetizing factor $N$ of the sample, we assume that the powder grains can be approximated as uniformly magnetized spheres occupying the cylindrical holder with volume fraction $\phi$ (estimated for this sample as $\phi \simeq 0.7$). We then refer to the approximation of randomly packed spheroidal particles \cite{Cas40,Ble41,Sko07,Bjo13,Bjo19} --- in particular, we write
\begin{equation}
	N = \left(1-\phi\right) N_{s} + \phi N_{c}
\end{equation}
where $N_{s} = 1/3$ and $N_{c} \simeq 0.1$ are the demagnetizing factors of a sphere and of the cylindrical holder in the considered geometry, respectively \cite{Cas40,Che06}. When analysing the isothermal curves as a function of the magnetic field, we first obtain the magnetization in volume units as
\begin{equation}
	M = \frac{m}{\phi V_{c}}
\end{equation}
where $m$ is the magnetic moment of the sample and $V_{c}$ is the volume of the cylindrical holder. Then, we plot $M$ as a function of the intrinsic magnetic field $H$ defined as
\begin{equation}
	H = H_{ext} - 4 \pi N M
\end{equation}
where $H_{ext}$ is the applied magnetic field.

\subsection*{Nuclear magnetic resonance}

We performed nuclear magnetic resonance experiments using an Apollo spectrometer from TecMag with a homemade setup. We selected {}$^{1}$H and {}$^{19}$F nuclear magnetic moments as local probes --- both are spin $I = 1/2$ nuclei and they are characterized by the gyromagnetic ratios $\gamma_{{}^{1}\rm{H}}/2\pi = 4.2576$ MHz/kOe and $\gamma_{{}^{19}\rm{F}}/2\pi = 4.0055$ MHz/kOe, respectively. Measurements were performed at the fixed magnetic field of $7.9$ kOe and $12.0$ kOe (generated using a Bruker electromagnet) and in the temperature range $5$ K $\leq T \leq 320$ K. To achieve these temperatures, we used static and dynamic flux cryostats with either liquid helium or liquid nitrogen as a cooling medium. The real temperature was estimated using either a thermocouple or a calibrated Cernox sensor (Lake Shore) located in close proximity to the sample. In order to minimize the possible effects of thermal history in the different cooling and warming cycles, we performed all the measurements upon cooling after a field-cooled protocol from above $300$ K.

Unrestrained powders of the sample were sealed in a quartz ampoule under an He atmosphere. The ampoule was inserted in a solenoidal coil which was tailored to maximize the geometrical filling factor. The coil as well as the variable capacitors were part of a resonant RLC circuit. After proper tuning and matching of the impedance at the working frequency, the coil was used both to generate the radiofrequency (RF) pulses and to detect the NMR signal inductively. Care was given to the choice of the coaxial cable along the probe using hydrogen-free and fluorine-free dielectrics for the measurements on {}$^{1}$H and {}$^{19}$F nuclei, respectively.

We measured the frequency-swept spectra and spin-lattice relaxation times at a fixed magnetic field. At fixed frequency, we cumulated the signal of a conventional Hahn-echo sequence of RF pulses over a fixed number of repetitions. The second half of the resulting spin-echo was fast-Fourier transformed and the integral $I$ of the real component of the resulting curve was associated to the corresponding frequency value. The overall spectrum $I(\nu)$ was obtained after repeating this procedure at different frequency values. The spin-lattice relaxation time $\textrm{T}_{1}$ of the sample was quantified by means of a conventional inversion-recovery sequence (with Hahn-echo reading and signal processing analogous to what described above for the frequency-swept spectra). The recovery of the nuclear magnetization towards thermodynamic equilibrium was obtained by plotting $I$ as a function of $\tau$, \textit{i.e.}, the waiting time between the inversion pulse and the Hahn-echo reading sequence. The experimental recovery curves for both {}$^{1}$H and {}$^{19}$F nuclei were consistent with single-exponential trends that can be fitted by the following expression
\begin{equation}\label{EqRecoveryFit}
	I(\tau) = I(\infty) \left\{1-2f\exp\left[-\left(\frac{\tau}{\textrm{T}_{1}}\right)^{\beta}\right]\right\}.
\end{equation}
Here, $\beta < 1$ is the so-called stretching parameter while the parameter $f \lesssim 1$ accounts for non-ideal inversion conditions. In both measurements, the RF pulses had typical duration around $1 - 2 \; \mu$s and the idle time $\tau_{r}$ between different repetitions was calibrated at all the temperature values so that $\tau_{r} \gtrsim 4 \; \textrm{T}_{1}$.

In general, $1/\textrm{T}_{1}$ is proportional to the spectral density $J(\omega_{L})$ of the time-modulated local magnetic fields fluctuating orthogonally to the quantization axis at the Larmor angular frequency $\omega_{L} = 2\pi\nu_{L}$ \cite{Abr61,Sli90}. According to the Bloembergen, Purcell, and Pound (BPP) model, the spectral density can be expressed in the Lorentzian form \cite{Blo48}
\begin{equation}\label{EqSpectralDensity}
	J(\omega) = \frac{\tau_{c}}{1+\omega^{2}\tau_{c}^{2}}
\end{equation}
if the autocorrelation function for the fluctuations is an exponentially-decaying function governed by the correlation time $\tau_{c}$. This leads to the formula
\begin{equation}\label{EqBPP}
	\textrm{T}_{1}^{-1}\left(T\right)^{\rm{BPP}} = C J(\omega_{L})
\end{equation}
where the factor $C \simeq \gamma^{2} \langle\Delta B^{2}\rangle$ is proportional to the gyromagnetic ratio $\gamma$ of the observed nucleus and to the mean square amplitude of the transverse fluctuating field $\Delta B$. The dependence of the spin-lattice relaxation rate on temperature is made explicit by assuming a specific dependence of the correlation time on temperature. For the Arrhenius-like slowing-down process typically described within the BPP model, in particular, $\tau_{c} = \tau_{0} \exp(\vartheta/T)$ where $\tau_{0}$ is the infinite-temperature asymptotic value of the correlation time and $\vartheta$ is the thermal equivalent of the activation energy.

\subsection*{Ferromagnetic resonance}

We performed ferromagnetic resonance experiments using a commercial Bruker EMX spectrometer equipped with an X-band resonator ER$4104$OR (TE$102$ mode, $\nu \simeq 9.46$ GHz). Under an inert Ar atmosphere, we mixed the powder sample with the two-component epoxy adhesive EpoFix (Struers) and glued it onto a quartz holder. This prevented the powder grains from being oriented by the magnetic field during the measurements and allowed us to insert the sample into the resonant cavity without exposing it directly to air and moisture. The temperature in the resonant cavity was controlled using a flow of nitrogen and monitored using a thermocouple located in the upper end of the cavity.

We measured the power ($P$) resonantly absorbed by the sample as a function of the applied magnetic field $H(t) = H + H_{a}(t)$ at fixed temperature. The quasistatic component $H$ was ramped from $300$ to $6300$ Oe at a rate of $\sim10$ Oe/s while the time-dependent field $H_{a}(t)$ ($|H_{a}| \simeq 0.5$ Oe) was modulated with a frequency of $100$ kHz. Using a lock-in detection at the modulation frequency, we measured the first derivative d$P$/d$H$ and then numerically integrated the signal to obtain $P$ as a function of $H$.

\section*{Data availability}

The data that support the findings of this study are available from the authors upon reasonable request.

\bibliography{Bibliography}

\section*{Acknowledgements}

We acknowledge Alessandro Lascialfari for stimulating discussions. We thank Alessandro Girella and Mauro Ricc\`{o} for their help with the sample handling and preparation under inert atmosphere. G.P. acknowledges support by the PNRR MUR project PE$0000023$-NQSTI. Synthesis was supported by the National Science Foundation through Award No. DMR-$2206534$. Preliminary magnetic measurements were supported by the Center for Molecular Quantum Transduction (CMQT), an Energy Frontier Research Center funded by the U.S. Department of Energy, Office of Science, Basic Energy Sciences, under Award No. DE-SC$0021314$. M.S.D. acknowledges graduate fellowship support from the NSF.






\end{document}